\documentstyle[12pt]{article}
\input html.sty
\begin{document}
\title{MATTERS OF GRAVITY, The newsletter of the APS TIG on Gravitation}
\begin{center}
{ \Large {\bf MATTERS OF GRAVITY}}\\
\bigskip
\hrule
\medskip
{The newsletter of the Topical Group in Gravitation of the American Physical 
Society}\\
\medskip
{\bf Number 10 \hfill Fall 1997}
\end{center}
\begin{flushleft}

\tableofcontents
\vfill
\section*{\noindent  Editor\hfill}

\medskip
Jorge Pullin\\
\smallskip
Center for Gravitational Physics and Geometry\\
The Pennsylvania State University\\
University Park, PA 16802-6300\\
Fax: (814)863-9608\\
Phone (814)863-9597\\
Internet: 
\htmladdnormallink{\protect {\tt pullin@phys.psu.edu}}
{mailto:pullin@phys.psu.edu}\\
WWW: \htmladdnormallink{\protect {\tt http://www.phys.psu.edu/PULLIN}}
{http://www.phys.psu.edu/PULLIN}
\begin{rawhtml}
<P>
<BR><HR><P>
\end{rawhtml}
\end{flushleft}
\pagebreak
\section*{Editorial}

Nothing profound to say in this editorial, just to thank the
contributors and correspondents that make this newsletter possible and
as usual to remind everyone that suggestions for authors/topics for
the newsletter are very welcome. I also want to apologize for the
delay in publication of this newsletter (it was due the first of the
month). It happened due to a major computer problem at our Center. In
the midst of the chaos and the rush to get the newsletter out as soon
as possible, some errors (additional to the usual ones...) might have
entered the newsletter. I apologize for those too.

The next newsletter is due February 1st.  If everything goes well this
newsletter should be available in the gr-qc Los Alamos archives under
number gr-qc/9709023. To retrieve it send email to 
\htmladdnormallink{gr-qc@xxx.lanl.gov}{mailto:gr-qc@xxx.lanl.gov}
(or 
\htmladdnormallink{gr-qc@babbage.sissa.it}{mailto:gr-qc@babbage.sissa.it} 
in Europe) with Subject: get 9709023
(numbers 2-8 are also available in gr-qc). All issues are available in the
WWW:\\\htmladdnormallink{\protect {\tt
http://vishnu.nirvana.phys.psu.edu/mog.html}}
{http://vishnu.nirvana.phys.psu.edu/mog.html}\\ 
A hardcopy of the newsletter is
distributed free of charge to the members of the APS
Topical Group on Gravitation. It is considered a lack of etiquette to
ask me to mail you hard copies of the newsletter unless you have
exhausted all your resources to get your copy otherwise.

If you have comments/questions/complaints about the newsletter email
me. Have fun.
\bigbreak

\hfill Jorge Pullin\vspace{-0.8cm}
\section*{Correspondents}
\begin{itemize}
\item John Friedman and Kip Thorne: Relativistic Astrophysics,
\item Raymond Laflamme: Quantum Cosmology and Related Topics
\item Gary Horowitz: Interface with Mathematical High Energy Physics and
String Theory
\item Richard Isaacson: News from NSF
\item Richard Matzner: Numerical Relativity
\item Abhay Ashtekar and Ted Newman: Mathematical Relativity
\item Bernie Schutz: News From Europe
\item Lee Smolin: Quantum Gravity
\item Cliff Will: Confrontation of Theory with Experiment
\item Peter Bender: Space Experiments
\item Riley Newman: Laboratory Experiments
\item Warren Johnson: Resonant Mass Gravitational Wave Detectors
\item Stan Whitcomb: LIGO Project
\end{itemize}
\vfill
\pagebreak

\section*{\centerline {April 1997 Joint APS/AAPT Meeting}}
\addtocontents{toc}{\protect\medskip}
\addtocontents{toc}{\bf News:}
\addtocontents{toc}{\protect\medskip}
\addcontentsline{toc}{subsubsection}{\it  April 1997 Joint APS/AAPT Meeting,
by Beverly Berger}
\begin{center}
    Beverly Berger, Oakland University\\
\htmladdnormallink{berger@vela.oakland.edu}
{mailto:berger@vela.oakland.edu}
\end{center}
\parindent=0pt
\parskip=5pt

For the second year, the Topical Group in Gravitation (GTG) has had a
significant presence at this meeting, which took place in Washington,
DC, April 18-21, with sponsorship of three invited sessions (two
jointly with other groups), two focus (special topics) sessions, and
two contributed sessions. The annual GTG business meeting was also
held. For those of you who were unable to attend but wish more
information than can be found in the summary given below, the
abstracts of most of the contributed and invited talks can be found at
\htmladdnormallink{http://www.aps.org/BAPSAPR97/}
{http://www.aps.org/BAPSAPR97/}.

The GTG co-sponsored with the Division of Particles and Fields an invited
session ``Frontiers of Theoretical Physics.'' Bob Wald spoke on cosmic
censorship. While this topic is normally of interest only to gravitational
theorists, you may recall that Kip Thorne's bet with Steven Hawking on this
subject made the front page of {\it The New York Times}. Wald
was able to capitalize on this excitement to present an excellent review of the
meaning of (primarily weak) cosmic censorship that  and whether known
counter-examples of naked singularities (including the Choptuik solution) were
generic. Again fortuitously,  Abhay Ashtekar was able to report on very recent
results in which geometrical operators in non-perturbative gravity could be
used to compute the quantum states of a black hole. This approach could then
immediately be compared with Juan Maldacena's discussion of the microscopic
calculation of black hole entropy in string theory using duality and D-branes.

The focus session on ``Analyzing Data from Gravitational Wave Detectors'' was
organized by Bill Hamilton. This represents an important area of interaction
between theorists and experimentalists: How to interpret the response of
current and future gravitational wave detectors. Sam Finn gave an invited talk
on data analysis for gravitational wave detectors emphasizing that one could
improve the statistical measure by considering all on-line detectors as a
single unit rather than considering each singlely and then looking for
coincidences. This new measure would be the probability or likelihood of the
combined signal profile. Warren Johnson then gave an invited talk on the
lessons learned with regard to data analysis from the Allegro bar detector at
LSU. He emphasized that all gravitational wave detectors to date have found
non-stationary, non-Gaussian noise sources whose level can be reduced but not
eliminated and that these should be treated as a ``background'' source of
signal in the data analysis. Even coincidences between two detectors may not be
sufficient to rule out such events. Other topics discussed in this session were
searching for burst gravitational waves using nonlinear filtering methods (E.
Flanagan), a pulsar search with Allegro (E. Mauceli), a rigorous way to
characterize observed coincidences in the absence of signal (A. Morse et al),
some novel ways to extend the frequency range of gravitational wave
interferometers (R. Drever), using the anelastic aftereffect to study thermal
noise in interferometer test masses (M.A. Beilby et al), and elimination of
some systematic error sources in Gravity Probe B (G.M. Kaiser et al).

Leonard Parker organized a focus session on black hole formation, evaporation
and entropy with several invited talks. Matt Choptuik described critical
phenomena in black hole formation. First discovered by Choptuik numerically,
the past few years have seen a growth of understanding of the nature of the
transition between initial data which collapse to a black hole and
those which disperse to infinity. Bob Wald, in an invited talk, argued that
the ``loss of information'' in the Hawking radiation process---pure state to
mixed state---was not a violation of quantum theory because the black hole
formation followed by evaporation creates a spacetime diagram that is not
equivalent to the one in which the black hole never existed. Ted Jacobson
continues to consider the issue of the role of field modes above the Planck
scale in the Hawking process. Standard derivations require such modes to be
present. However, Jacobson reports on calculations showing that it is possible
for energy that would in principle come from such modes to appear at lower
energies (mode conversion) without seriously altering the thermal spectrum of
black hole evaporation. In his invited talk, Larry Ford discussed the role of
quantum fluctuations in the stability of black hole horizons. He showed that
while quantum effects could perturb the horizon, the effect on the Hawking
radiation is small for black holes with masses above the Planck mass. Leonard
Parker, meanwhile, showed that in his exactly solvable $1 + 1$ dilaton gravity
model, there is a threshold mass for black hole formation (in contrast to the
$3 + 1$ Choptuik result). Contributions by Eric Martinez on a thermodynamic
formalism that incorporates strong gravitational fields and a discussion by
David Brown on the role of boundary states in black hole entropy completed this
session.

The joint invited session between GTG and the Topical Group on Fundamental
Constants and Precision Measurements was again very successful. This time, the
emphasis was on ``Sensitive Mechanical Measurements and the Detection of
Gravitational Radiation.'' Peter Saulson led off with an overview of how to
detect a feeble signal amidst the noise. He emphasized lessons learned from Bob
Dicke---perform a null measurement and use modulation to enhance the effect of
the signal. Jennifer Logan then described the efforts made over the past
several years to reduce the mechanical noise in the LIGO 40 meter detector. She
also discussed the use of this prototype in the development of power recycling
and other advanced LIGO techniques. Bill Hamilton then gave a talk in which he
reviewed the development and progress of the Allegro detector. This device has
operated almost continuously for the past 5 years and has given great insight
into noise reduction and the problems associated with continuous operation. He
also discussed proposed spherical detectors. Finally, Mark Bocko reviewed the
state of the art in quantum non-demolition techniques and how they might be
applied to the detection of weak forces.

The GTG invited session was entitled ``Sources and Detection of Gravitational
Waves.'' Jorge Pullin discussed the ``close approximation''---the treatment of
two black holes as a single distorted black hole. In second order perturbation
theory, the accuracy of the approximation can be studied. Perhaps surprisingly,
the analytic results agree quite well with numerical results---for separations
larger than one would expect.  Bruce Allen reported on calculations of how a
stochastic background of gravitational waves might be detectable. He considered
the sensitivity of LIGO to such a background. Joan Centrella described recent
work on numerical simulations of infalling neutron star binaries and the
gravitational radiation they produce. Smooth particle hydrodynamics with
Newtonian gravity was used to describe the neutron stars while the quadrupole
approximation was used to calculate the gravitational radiation. These
approximations allowed many simulations to be run in order to study the
dependence of the waveforms and spectra on the parameters of the binary and
infall. Finally, David Shoemaker reported on the status of the LIGO
construction. The highlights of his talks were photographs of one completed arm
at the Hanford site and of the construction progress at both sites. He also
described some of the nuts and bolts issues of operation and the development of
the laboratory and users group. 

There were also two sessions of contributed papers. Highlights included a
series of experimental talks on the proposed space-based LISA interferometer
(R.T. Stebbins et al), noise reduction using active vibration isolation (J.A.
Giaime et al and S.J. Richman et al), a balanced heterodyne detection scheme
for signal extraction (K.-X. Sun),  and using VLBI for solar system gravitation
tests (T.M Eubanks et al). There were also several talks on numerical
simulations of close compact binaries (New and Tohline), inspiraling neutron
star binaries (Matthews and Marronetti), critical phenomena in a harmonic map
model (Liebling and Choptuik), and velocity dominance in Gowdy cosmologies
(Berger and Garfinkle). In addition, there were analytic discussions on a
prescription for relativistic quantization (C. Vuille) and Cauchy horizon
stability in plane-wave spacetimes (Konkowski and Helliwell). Several other
experimental and theoretical talks were also given.

Kip Thorne chaired the GTG business meeting which was followed by the
business meeting of the LIGO Research Community chaired by Sam Finn.

\vfill
\pagebreak
\section*{\centerline {The Physics Survey}\\\centerline{and the Committee on 
Gravitational Physics}}
\addtocontents{toc}{\protect\medskip}
\addcontentsline{toc}{subsubsection}{\it The physics survey and committee on 
gravitational physics,
by Jim Hartle}
\begin{center}
    Jim Hartle,  UC Santa Barbara\\
\htmladdnormallink{hartle@cosmic.physics.ucsb.edu}
{mailto:hartle@cosmic.physics.ucsb.edu}
\end{center}
\parindent=0pt
\parskip=5pt

Once a decade, the National Research Council's Board on Physics 
and Astronomy (BPA) conducts a survey of the fields of physics. The most
recent was the eight volume  {\sl Physics Through the 1990's}
in 1986, familiarly known as the ``Brinkman Report'' after the chair of the 
survey committee. This was preceded by the ``Bromley Report'' in 1972. 
The BPA is now carrying out a new decadal survey entitled {\sl Physics
in a New Era} under a committee chaired by Dave Schramm. 

These surveys play an important role in conveying the consensus
of the scientific community on past achievements and the future priorities 
to decision makers in Washington, both in funding agencies
and the Congress. They are also an opportunity to strengthen
the understanding of physics generally and foster its support. 
It seems likely that the new survey will be particularly important
as it comes  at the start of a period of constrained funding for science.

\bigskip
{\it  The Committee on Gravitational Physics}

The new survey will include a volume on each of the major branches
of physics, as well as an overview volume. 
For the first time there will be a separate volume on 
gravitational physics prepared by a Committee on Gravitational
Physics (CGP). The members of CGP are:

James B. Hartle, Chair, University of California, Santa Barbara\hfil\break
Eric G. Adelberger, University of Washington\hfil\break
Abhay V. Ashtekar, Pennsylvania State University\hfil\break
Beverly K. Berger, Oakland University\hfil\break
Gary T. Horowitz, University of California, Santa Barbara\hfil\break
Peter F. Michelson, Stanford University\hfil\break
Ramesh Narayan, Harvard-Smithsonian Center for Astrophysics\hfil\break
Peter R. Saulson, Syracuse University\hfil\break
Joseph H. Taylor, Jr., Princeton University\hfil\break
Saul A. Teukolsky, Cornell University\hfil\break
Clifford M. Will, Washington University

The first meeting of the CGP will be in Washington on October 7-9, 1997. 
The committee hopes to have finished its task by summer, 1998.              

The objectives of the report are as follows:

{$\bullet$} Describe the progress in gravitational physics
in the last decade. 

{$\bullet$} Identify the scientifically promising directions
for the next decade, and describe the experimental, observational, 
and theoretical resources that are required to pursue these directions. 

{$\bullet$} Describe the relationships of gravitational physics
to neighboring areas of science, in particular, astrophysics, particle
physics, cosmology, and mathematics. 

{$\bullet$} Assess the standing of the US effort in gravitational
physics relative to that in other countries and identify opportunities
for international collaboration. 

{$\bullet$} Examine career patterns and opportunities for scientists
in gravitational physics and assess the implications of these for 
the support of students, post-doctoral researchers, and faculty. 

\bigskip
{\it Input to the Committee}

The committee invites input from scientists working in gravitational
physics that are related to the above objectives, and the individual
members of the committee would be pleased to discuss such input. 
Input should be sent to the chair by e-mail at \hfil\break
{\tt hartle@cosmic.physics.ucsb.edu} or by letter at: 

{James B. Hartle}
{Department of Physics}
{University of California}
{Santa Barbara, CA 93106}

As an aid to focusing input, 
the following are some of the kinds of questions that the CGP will
be seeking answers to. This list is not meant as an
opinion poll, and it is not expected that every input will
address all questions. What would be most helpful are brief, reasoned arguments
supporting definite directions in research and funding. Responses
like ``X-theory should have the highest priority, Sincerely, Prof. Z ''
are therefore not helpful. On the other hand, copies of your grant 
proposals are probably too long and too specific. Please try to be 
realistic. It is commonly agreed that we are facing an era of constrained
support for science, and the best that can be hoped for for the NSF budget
is level funding. Even if such projections prove overly
pessimistic, it is better to be prepared for underfunding rather
than the reverse. Responses concluding that the funding for theoretical
gravity should be tripled or  that we should construct  accelerators
at Planck energies are also not helpful. 

1. What, in your view, are the most outstanding achievements
in gravitational physics in the past decade?

2. What, in your opinion, are the most promising directions  
for research in gravitational physics in the next decade?

3. What resources -- in people and facilities -- are needed
to realize these opportunities?

4. What are the most persuasive arguments that the nation should
allocate these resources in competition with other opportunities 
in science?

5. What should be the top priorities in the NSF program on
gravitational physics, and what should be the lowest priorities, 
assuming a level or declining budget? 

6. Large facilities or projects are becoming increasingly important
in some areas of gravitational physics -- GPB and LIGO for example. 
Large facilities like LISA and STEP are proposed. How important
are these projects for the progress in gravitational physics and
which are the most important?

7. What should be done by funding agencies to improve
career opportunities for gravitational scientists, and how
important are these improvements compared to preserving 
the existing core research program?

8. How does the US effort in gravitational physics compare
with that in other countries? What are the implications
of international competition in the area and what are the desirable
opportunities for international collaboration?

9. Theoretical progress in some areas has come to increasingly
depend on large computer simulations that require collaborations
of many scientists. How important are these efforts, what
are the resources required,  and what is the best way to organize 
these efforts. 

10. Is there adequate theoretical support for the prediction and
analysis of presently planned and future experiments?

11. To what problems in astrophysics,  cosmology, and high
energy physics  can gravitational physics contribute to and what
areas of gravitational physics research, both theoretical and
experimental, should be emphasized from this point of view?

12. Should we foster greater cooperation and interaction between
high energy theorists and gravitational theorists working
on fundamental questions in quantum gravity? If so what is the
best way to achieve this? 

13. What is your view on the role research in gravitational
physics plays in the education of people who go on to do useful
things outside the field?

14. What other issues concerning the future of gravitational
physics should the CGP address, in your opinion? 

\bigskip
{\it Further Information}

The BPA website: 
\htmladdnormallink{http://www.nas.edu/bpa}{http://www.nas.edu/bpa}\hfil\break
 The Physics Survey website: 
\htmladdnormallink{http://www.nas.edu/physsurv.html}
{http://www.nas.edu/physsurv.html}\hfil\break
 The CGP website: 
\htmladdnormallink{http://www.nas.edu/cgp.html}{http://www.nas.edu/cgp.html}

\vfill\pagebreak
\section*{\centerline {Instability of rotating stars to axial perturbations}}
\addtocontents{toc}{\protect\medskip}
\addtocontents{toc}{\bf Research briefs:}
\addtocontents{toc}{\protect\medskip}
\addcontentsline{toc}{subsubsection}{\it 
Instability of rotating stars to axial perturbations, by Sharon Morsink}
\begin{center}
    Sharon Morsink,  University of Wisconsin, Milwaukee\\
\htmladdnormallink{morsink@csd.uwm.edu}{mailto:morsink@csd.uwm.edu}
\end{center}
\parindent=0pt
\parskip=5pt

A great wealth of information about the internal structure of neutron
stars can be obtained through the study of neutron star
oscillations. Just as helioseismology has recently revealed important
details of the sun's structure, it may be possible in the fu ture to
detect gravitational waves caused by the oscillations of a relativistic
star and obtain the star's mass and radius [1]. The possibility
of making such exciting measurements underlines the importance of
understanding the theoretical details o f the pulsations of neutron
stars. Earlier this year Nils Andersson [2] made a
surprising discovery while numerically investigating the subset of
non-axisymmetric perfect fluid oscillations known as axial
perturbations: all axial perturbation s with azimuthal angular
dependence $e^{im\phi}$ are unstable when the star rotates, for any
value of the star's angular velocity.  As a result, all rotating stars
are unstable to small perturbations!

It has been known for some time that rotating stars are unstable to
gravitational radiation reaction [3,4,5] via the
Chandrasekhar-Friedman-Schutz (CFS) instability. It turns out that
Andersson's result can be explained by the CFS mechanism, but the way
that the instability sets in is different from the usual result which
holds for polar perturbations. (Recall that the non-radial fluid
velocity field created by a polar perturbation of a spherical star can
be expressed as a gradient of a spherical harmonic, while for an axial
perturbation it is a cross-product of a radial vector and a polar
flow.) For a polar perturbation with fixed value of $m$, the
perturbation is stable for small stellar angular velocity, $\Omega$,
until a critical velocity, $\Omega_c$ is reached. When
$\Omega=\Omega_c$, the perturbation's frequency vanishes as seen by
inertial observers. For all angular velocities $\Omega >\Omega_c$, the
mode is unstable. Andersson's result is that axial modes are unstable
for all $
\Omega>0$. 

The difference in critical velocities for the two types of
perturbations is really not too surprising. For static stars, axial
fluid perturbations are trivial [6] and their frequencies of
oscillation must vanish [7]. This implies that the critical
angular velocity is zero and that axial modes are unstable for any
non-zero angular velocity. Indeed, Papaloizou and Pringle [8]
have studied these modes for Newtonian stars (which they call r-modes)
and the form of the frequency which t hey calculate conforms to the
CFS instability criterion. However, the implied instability of the
Newtonian r-modes went unnoticed until Andersson pointed it out
[2]. A formal proof of the instability for the general
relativistic analogue of t he r-modes in the slow rotation limit is
presented in a paper by John Friedman and me [7].

Of what astrophysical significance is this new instability? If the
 instability's growth time is shorter than the time scale for
 viscosity to damp it out, the axial mode could be an important source
 of gravitational radiation. For a $l=m=2$ axial mode the
 instability's growth rate scales as $(\Omega\sqrt{R^3/M})^{10}$ (in
 geometrical units, where $\Omega\sqrt{R^3/M}\ll 1$) while the damping
 rate due to shear viscosity is independent of $\Omega$.  An order of
 magnitude calculation (which agrees with preliminary numerical
 results [9]) shows that for a neutron star with a
 temperature of $10^9 K$, the two time scales are equal when the
 rotational period is of the order of a millisecond. (Assuming a
 coefficient of shear viscosity which takes account of superfluid
 effects [10].) As the star rapidly cools, shear viscosity will
 increase and quickly damp out the instability.  This leaves open
 some interesting questions for future research.  When viscosity is
 included in a full relativistic computation, do axial or polar
 perturbations place the lower limit on the angular velocity of 
neutron stars born with high angular momentum? As the newborn star
 cools and spins down, is it possible for the star to be in a
 marginally unstable configuration for a long enough time so that an
 appreciable amount of gravitational radiation is emitted? We look
 forward to the resolution of these problems.

{\bf References:}

[1] N. Andersson and K.D. Kokkotas, 
\htmladdnormallink{gr-qc/9610035}{http://xxx.lanl.gov/abs/gr-qc/9610035}
\hfil\break
[2] N. Andersson, \htmladdnormallink{gr-qc/9706075}
{http://xxx.lanl.gov/abs/gr-qc/9706075}.\hfil\break
[3] S. Chandrasekhar, Phys. Rev. Lett., {\bf 24}, 611 (1970).\hfil\break
[4] J.L. Friedman and B.F. Schutz, ApJ, {bf 222}, 281 (1978).\hfil\break
[5] J.L. Friedman, Commun. Math. Phys., {\bf 62} 247 (1978).\hfil\break
[6] K.S. Thorne and A. Campolattaro, ApJ, {\bf 149} 591 (1967).\hfil\break
[7] J.L. Friedman and S.M. Morsink, \htmladdnormallink{gr-qc/9706073}
{http://xxx.lanl.gov/abs/gr-qc/9706073}.\hfil\break
[8] J. Papaloizou and J.E. Pringle, MNRAS, {\bf 182} 423 (1978).
\hfil\break
[9] N. Andersson, personal communication.\hfil\break
[10] C. Cutler and L. Lindblom, ApJ, {\bf 314} 234 (1987). 
\vfill
\pagebreak
\section*{\centerline {LIGO project status}}
\addtocontents{toc}{\protect\medskip}
\addcontentsline{toc}{subsubsection}{\it 
LIGO project status, by Stan Whitcomb}
\begin{center}
    Stan Whitcomb, Caltech\\
\htmladdnormallink{stan@ligo.caltech.edu}{stan@ligo.caltech.edu}
\end{center}
\parindent=0pt
\parskip=5pt

Construction continues to move forward rapidly at both LIGO sites
(Hanford, Washington and Livingston, Louisiana).  At the Hanford site,
the civil construction at the site (buildings, roads, power) is nearing
completion.  At the Livingston site, the main activities are the
construction of the buildings and the forming of the concrete
foundation along the two arms on which the beam tubes will be
installed.

The vacuum system is also moving forward.  Chicago Bridge and Iron, the
company building the LIGO beam tubes (which connect the vertex and ends
of the two arms), has completed the fabrication and installation of all
8 km of  beam tube at the Hanford site. The first two 2 km sections have
been evacuated and are already at a pressure below $ 10^{-6}$ torr.
They have now moved their fabrication equipment to a facility near the
Livingston site, and are starting to prepare for full production.  Our
contractor for the fabrication of the vacuum chambers and associated
equipment which will be in the located in the buildings, Process
Systems International, is nearing completion of all the large chambers
and associated hardware for the Hanford site.  Installation
is expected to start in September.

The staffing of the sites is also starting; approximately 15 LIGO staff
are located at the two sites, including Hanford Site Head Fred Raab,
who recently moved there from Caltech.

The design of the LIGO detectors is accelerating, with the various
detector subsystems split approximately 50-50 between the preliminary
and final design phases.  Fabrication has started for long-lead items
including the test masses and other large optics.  Approximately half
of the fused silica blanks have been received with the remainder
expected before the end of the year; General Optics and the
Commonwealth Scientific and Industrial Research Organization are
polishing these blanks in preparation for coating.  Procurements are
underway for a complete first article Seismic Isolation Stack to be
built this fall with testing to start in the beginning of 1998.

Lightwave Electronics Corporation, under contract to develop a 10 watt
single frequency Nd:YAG laser for LIGO, has completed the design and
are starting fabrication of the first unit.  An experimental unit used
to test the performance of this new design met the key requirements for
power, beam quality, frequency and intensity noise.

At MIT, a 5 m long suspended interferometer is being used to
investigate the limits of optical phase measurements. This recycled
Michelson interferometer operated initially with an Argon ion laser at
514 nm, and demonstrated a sensitivity of $ 3 \times 10^{-10}$ rad
Hz$^{-1/2}$.   It has now has now been converted to operate with a
Nd:YAG laser at 1064 nm.   A detailed characterization of the noise in
this new configuration will begin soon.

The initial meeting of the LIGO Scientific Collaboration (LSC) was
held in Baton Rouge, Louisiana in August.  The purpose of this meeting
was to form a broader scientific effort to both develop the initial
detectors and to pursue research leading to more sensitive future
detectors.  Twenty groups from five countries, representing a total of
201 collaborators were represented.  The most important agenda items
were to discuss a charter for the LSC and to form working groups on
specific technical topics to coordinate the research efforts of
different groups.  Rai Weiss (MIT) was appointed as the first
spokesperson for the collaboration.  The next meeting of the LSC is 
scheduled for March 12-13 at the LIGO Hanford site.

Additional information about LIGO, including our monthly newsletter and
information about the LSC, can be accessed through our WWW home page at
\htmladdnormallink{http://www.ligo.caltech.edu}{http://www.ligo.caltech.edu}.
\vfill\pagebreak
\section*{\centerline {The Search for Frame-Dragging }}
\addtocontents{toc}{\protect\medskip}
\addcontentsline{toc}{subsubsection}{\it 
{The Search for Frame-Dragging, by Clifford Will}}
\begin{center}
    Clifford Will, Washington University, Saint Louis\\
\htmladdnormallink{cmw@howdy.wustl.edu}{cmw@howdy.wustl.edu}
\end{center}
\parindent=0pt
\parskip=5pt
\section*
\noindent

Gravitomagnetism has a history that is at least as long as that of
general relativity itself.  The idea that mass currents might generate
the gravitational analogue of magnetic fields, and crude experiments
to look for such effects predated Einstein.  Soon after the
publication of general relativity (GR), Lense and Thirring calculated
the advance of the pericenter and line of nodes of a particle orbiting
a rotating mass.

The gravitomagnetic ``dragging of inertial frames'' by rotating matter
has played a part in discussions about the meaning and usefulness of
Mach's principle, in astrophysical models of jets near accreting,
rotating black holes, and in proposals for testing alternative
theories of gravity.

It is no surprise then, that substantial effort during the past 30 or
so years has gone into trying to measure gravitomagnetism.  A recent
preprint by Ignazio Ciufolini and colleagues [1] claims to have
succeeded.

There are three main effects of gravitomagnetism in the solar-system:

1. {\it Precession of a gyroscope}. \quad In the field of a body with angular
momentum $\vec J$, a gyroscope at a distance $r$ 
precesses with an angular velocity given 
by
${\vec \Omega}_{\rm gyro} = -\mu ({\vec J}-3{\vec n} {\vec n} 
\cdot {\vec J})/r^3\,,$
($G=c=1$) where $\mu$ denotes the coefficient of frame dragging (1
in GR, ${1 \over 2} (1+\gamma+ \alpha_1/4)$ in the PPN framework). 
For a gyroscope in a polar
Earth orbit at 600 km altitude, the rate is 43 milliarcseconds
(mas) per year.   

2. {\it Precession of orbital planes}. \quad 
The orbit of a particle 
is a ``gyroscope'', whose axis or ``node'' (intersection of the orbit
with a reference plane) will also precess.  
The rate is given by
$\vec \Omega_{\rm node} = 2 \mu {\vec J} /a^3 (1-e^2)^{3/2} \,,$
where $a$ and $e$ are the semi-major axis and eccentricity of the
orbit.
For a satellite at 5000 km altitude, it amounts to about 31
mas per year.  

3. {\it Precession of the pericenter}. \quad In the field of a rotating body 
there is an advance of
$\dot \omega_{\rm pericenter} = -4 \mu |\vec J| \cos I /a^3 (1-e^2)^{3/2}
\,,$
where $I$ is the orbital inclination.

Since the early 1960's, measurement of the first effect has been the
goal of the Stanford Gyroscope experiment (Gravity
Probe B). 
The goal
is to measure the precession of an array of gyroscopes in low
Earth orbit to better than one percent.  
Following years of
financial uncertainty, the project was
endorsed in 1995 by a panel convened by the National Academy of
Sciences [2],
and NASA Administrator Daniel Goldin made a firm commitment to the mission. 
The
spacecraft and payload are under construction at Stanford and
Lockheed-Martin, and the project is actually slightly ahead of
schedule for launch in December 1999 [3].

The paper by Ciufolini {\it et al.} is based on measuring
the second effect, the nodal precession.    The original idea was
proposed in the late 1950s by Husein Yilmaz, and later embellished 
by Richard Van Patten and Francis Everitt: measure the
precession of the plane of a satellite in polar orbit.
The multipole moments of the
Earth's gravitational field also induce orbital precession via standard
Newtonian gravity, but for polar orbits, the effects
vanish.  It's crucial to suppress the Newtonian effects,
because they amount to about $368 ~\cos I$ degrees per year.  
(At 12 degrees inclination, the precession is 360 degrees
per year, permitting sun-synchronous orbits.)  

Ciufolini proposed a generalization of the Yilmaz-Van Patten-Everitt
idea.  Since the effect of the even-order Newtonian multipoles is
proportional to $\cos I \times$ functions of $\cos^2 I$, one can
cancel the Newtonian effects using two satellites in orbits whose
inclinations are supplementary ($I_1+I_2=180^{\rm o}$).  (The Earth's
odd-order multipoles, $L=3,5 \dots$ are not important).  He then noted
that there already existed one satellite for this purpose: the Laser
Geodynamic Satellite (LAGEOS), a massive, 60 cm diameter sphere,
studded with laser retro-reflectors, which was launched into a nearly
circular orbit with $I \approx 110^{\rm o}$ in 1976, and soon became a
central tool in geophysics and geodynamics.  Low atmospheric drag, and
the centimeter accuracy of laser ranging were key to its usefulness.

All that was needed for a frame-dragging test at around a 10 percent
level was a LAGEOS in an orbit of $70^{\rm o}$ inclination.  Alas,
this was not to be, and when LAGEOS II was launched in 1992,
geophysical and political criteria dictated $I=53^{\rm o}$.  Although
Ciufolini and others lobbied hard for a LAGEOS III with a suitable
inclination, it has not yet materialized.

Nevertheless, Ciufolini and co-workers have argued that the situation
is not hopeless.  The Earth's multipole moments are known very
accurately, from decades of accurate measurements of satellite orbits
(including LAGEOS).  Moments $J_6$ and higher are small enough and are
known well enough that their effects can be subtracted off.
Unfortunately, $J_2$ and $J_4$ are not known quite well enough.  Thus
the effective {\it measured} nodal precession can be viewed as a
linear combination $\Omega_{\rm node}^{\rm obs}=A(I) \Delta J_2+B(I)
\Delta J_4+C\mu \,,$ where $\Delta J_i$ denote the errors in $J_2$ and
$J_4$ and $\mu$ is the frame-dragging coefficient to be measured.
Thus there are two measurables, but three unknowns -- it's only in the
supplementary inclination case that the $J_2 ~-~ J_4$ linear
combinations are degenerate, and $\mu$ can be determined uniquely with
only two observables.  Given the two non-supplementary LAGEOS
satellites, one needs a third measurable.  By happy chance, LAGEOS II
turned out to have a decent eccentricity -- 0.014, as compared to
0.004 for LAGEOS I.  This makes its perigee advance measurable.  But
the predicted advance has a different dependence on the Earth's
moments and on frame-dragging: $\dot \omega_{\rm pericenter}^{\rm
obs}=A^\prime (I) \Delta J_2+ B^\prime (I) \Delta J_4+C^\prime \cos I
\mu $.  According to Ciufolini {\it et al.}, this gives the third
measurable needed.

But this quantity is the weak link in the chain for several reasons.
First, the measured orbital displacements are proportional to $e \dot
\omega$ and $e$ is still pretty small, so while the nodal precessions
could be measured to 1 mas per year, the pericenter advance was
limited to about 10 mas per year accuracy.  Second, the effects of the
odd-order moments are significant for the pericenter advance.  Third,
non-gravitational perturbations of the satellite, such as those
related to radiation pressure and thermal heating, affect the
pericenter advance more strongly than they do the nodal advance.
Also, tidal, secular, and seasonal variations in all the moments must
be carefully taken into account in both nodal and pericenter
precessions.  The reported result for $\mu$ was 1.1, with a realistic
error of about 25 percent ($\mu_{GR}=1$).  By contrast, researchers at
the University of Texas argue that, in view of the many error sources,
an error of 200 percent is probably more realistic [4].

As in all such satellite experiments, with many corrections to be made
and subtle systematic effects to be dealt with, more data and an
independent data analysis are called for to see if a LAGEOS I \& II
experiment can really detect gravitomagnetism.  In any case, the NASA
relativity mission should be much higher precision (by a factor at
least 25), thought admittedly at a much higher price tag.

{\bf References}

[1] I. Ciufolini, D. Lucchesi, F. Vespe and F. Chieppa,
\htmladdnormallink{gr-qc/9704065}{http://xxx.lanl.gov/abs/gr-qc/9704065}, 
submitted to {\it Nature}\hfil\break
[2]  Truth in advertising: panel which included the present
correspondent.\hfil\break
[3]  See 
\htmladdnormallink{http://stugyro.stanford.edu/RELATIVITY}
{http://stugyro.stanford.edu/RELATIVITY}\hfil\break
[4]  J. Ries, private communication.
\vfill\pagebreak
\section*{\centerline {Conference of the Southern African Relativity 
Society}}
\addtocontents{toc}{\protect\medskip}
\addtocontents{toc}{\bf Conference reports:}
\addtocontents{toc}{\protect\medskip}
\addcontentsline{toc}{subsubsection}{\it 
Conference of the Southern African Relativity Society, by Nigel Bishop}
\begin{center}
    N T Bishop, University of South Africa, Pretoria\\
\htmladdnormallink{bishont@alpha.unisa.ac.za}{bishont@alpha.unisa.ac.za}
\end{center}
\parindent=0pt
\parskip=5pt

The Southern African Relativity Society was founded in 1995 at a meeting at
the University of Zululand. This, the second conference of the society,
was held at the University of South Africa, Pretoria, on 6 and 7 February
1997. The conference was organized by the Council of the society (G.F.R.
Ellis (President), A. Beesham, N.T. Bishop, W. Lesame and S.D. Maharaj),
with local organizing committee consisting of N.T. Bishop, F.E.S. Bullock
and S.D. Maharaj. The conference was funded by the University of South
Africa and the Foundation for Research Development.

There were 31 delegates at the conference: mainly from South Africa, but
also from Egypt, India, Italy, Malawi, Nigeria, Russia, U.K. and U.S.A.
The plenary speakers were R.A. Isaacson (N.S.F., U.S.A.), J.V. Narlikar
(I.U.C.A.A., India) and J. Winicour (Pittsburgh, U.S.A.).

Richard Isaacson reported on the LIGO project, which is expected to open a
new window on the Universe in about 2001; of course, the interesting things
that will be seen through this window are those that are not anticipated.
Jayant Narlikar talked about the revival of the cosmological constant,
arguing that the standard FRW model does not satisfy the observational
constraints imposed by the ages of globular clusters, etc. Jeffrey Winicour
discussed the optics of black hole formation, and showed computations of
the caustics of the event horizon in the axisymmetric case.

Research in relativity in South Africa is concentrated at three centres:
Cape Town, Durban and Pretoria. The best known group is probably that at
Cape Town led by George Ellis. Their work is now very much focussed on
cosmology, and includes the cosmic microwave background, gravitational
lensing, observational cosmology and almost-FRW universes. There are more
relativists in and around Durban than in Cape Town, not because there is
one large group in Durban, but because there are several universities in
the area each with an active interest in relativity. Their interests
include symmetries and exact solutions, cosmology and inflation, and
computer algebra. The group in Pretoria (led by Nigel Bishop) mainly works
on numerical relativity, in collaboration with the Binary Black Hole
Alliance in the U.S.A. Other interests include observational cosmology,
computer algebra and numerical analysis.

In conclusion, the conference provided a useful opportunity for
discussion amongst relativists in southern Africa and other parts of
the world. The next conference is scheduled for early 1999 in Cape Town.
The Conference Proceedings (participants and abstracts) are available on
the world wide web at:
\htmladdnormallink{http://shiva.mth.uct.ac.za/SARS/}
{http://shiva.mth.uct.ac.za/SARS/}
\vfill\pagebreak

\section*{\centerline {II Warszaw workshop on canonical and quantum gravity}}
\addtocontents{toc}{\protect\medskip}
\addcontentsline{toc}{subsubsection}{\it 
{II Warszaw workshop on canonical and quantum gravity, by Carlo Rovelli}}
\begin{center}
    Carlo Rovelli, University of Pittsburgh\\
\htmladdnormallink{rovelli@phyast.pitt.edu}{rovelli@phyast.pitt.edu}
\end{center}
\parindent=0pt
\parskip=5pt

The canonical quantum gravity community gathered in Warsaw, in June, 
for the second, but already traditional, Warsaw workshop.  A perfect 
organization, mostly the merit of Jerzy Lewandowski, and a well chosen 
balance between scientific focus and openings towards nearby areas, 
have contributed to a dense and exciting meeting.  The field is in 
fibrillation, with excitement, new ideas, and a feel of progress 
happening; the talks were all packed; and the discussion lively.

By far the largest topic discussed (half of the talks), has been 
{\it loop quantum gravity}.  On the side of physical results, 
Kirill Krasnov and Abhay Ashtekar reported substantial progress on the 
problem of deriving {\it black hole entropy} from quantum gravity, 
developing the earlier works on the subject by Krasnov and Rovelli.  
Surprisingly, the long searched derivation of the black hole entropy 
formula from quantum gravity has being found, almost simultaneously, 
in both the current major approaches to quantum gravity: strings and 
loop gravity.  The two derivations have opposite strengths and 
weaknesses.  The string theory one succeeds in computing the precise 
entropy/area ratio ($1/4\hbar G$), but so far it works only for highly 
unphysical (extreme or nearly extreme) holes; while the loop 
derivation works for physical cases such as Schwarzschild, but it 
does not fix the $1/4\hbar G$ factor (although it is compatible with 
it).

Four talks were devoted to a novel direction in loop quantum gravity: 
spacetime covariant versions of the formalism.  Mike Reisenberger and 
Carlo Rovelli showed how one can derive a sum-over-histories  
formulation of loop quantum gravity from the canonical theory, 
following earlier ideas by Reisenberger himself and Baez.  The 
resulting theory has the intriguing form of {\it a sum over 
topologically inequivalent surfaces in spacetime}.  Fotini Markopoulou 
and Lee Smolin explored Lorentzian versions of this construction.  
Covariant formalism do not seem to be an appropriate topic for a 
workshop on {\it canonical} gravity!  But maybe old antinomies as 
the 4 versus 3+1 views of quantum gravity are finally beginning to 
evaporate.

The weak side of loop quantum gravity is the dynamics, still much 
debated.  Thomas Thiemann, who has recently given a key contribution 
by constructing a well-defined hamiltonian operator, described the 
extension of his results to the {\it inclusion of matter}.  The 
attractive aspect of this new step is that finiteness of the matter 
hamiltonian supports the hope that loop quantum gravity could realize 
the dream of curing ultraviolet divergences.  The discussion on the 
physical correctness of the proposed hamiltonian and its variants 
focused on the problem of the existence of {\it anomalies} in the 
constraint algebra.  Roman Jackiw emphasized the importance of the 
problem by discussing some model theories.  Don Marolf reported on an 
elegant and comprehensive analysis of the constraint algebra by 
Lewandowski and himself: The algebra closes in most of the proposed 
versions of the hamiltonian constraint.  However, it does not seem to 
reproduce the classical algebra, and doubts were thus raised on the 
physical correctness of the proposed operators. 

Other aspects of loop quantum gravity were discussed by Jorge Pullin 
in a comprehensive review of the state of the Chern-Simon state in the 
theory, including recent results obtained using the spin-network 
technology, and by Renate Loll, who introduced a novel technique for 
the computation of the spectrum of the volume.

The second largest topic discussed, after loop quantum gravity, has 
been the problem of formulating quantum mechanics in a form 
appropriate for gravity.  Jim Hartle reviewed his {\it generalized 
quantum theory} emphasizing its numerous applications.  Chris Isham 
discussed the formalism of consistent histories, focusing on the 
intriguing appearance of {\it topos theory}, a sophisticated branch 
of mathematics, at its roots.  The difference in mathematical style 
did not hide the fact that these two speakers were talking about the same 
formalism.  A formalism which has become very relevant for loop quantum 
gravity, in view of the recent steps towards spacetime, 
sum-over-surfaces  
formulations, which fit naturally into the Hartle-Isham quantum 
mechanics.  Chris left Warsaw before giving his final lecture, 
due to an indisposition; but on my way back from Warsaw I had 
the fortune of spending a delightful day with him in London, and I 
can assure anybody who might have worried that he was again in 
perfect conditions! 

Ted Newman illustrated the intriguing {\it reformulation of 
general relativity in terms of null surfaces}, recently completed 
by himself, Frittelli, Kozameh and others, including applications 
to the quantum theory.  Mauro Carfora described his analytical 
derivation of the existence and location of a critical point in 
{\it simplicial quantum gravity}.  Abhay Ashtekar presented  
some puzzling ``large'' quantum gravitational effects.  Peter 
Hajicek discussed the quantization of $2+1$ gravity.  

Other subjects covered, which I can only list here for lack of 
space, were quantum field theory on curved spacetime 
(Fredenhagen); $\theta$ angles (Landsman); relativistic 
hydrodynamics (Kijowski); exterior gauge fields (Henneaux); dust 
(Kuchar) and spherical dust shells (Louko); the canonical 
structure of homogeneous cosmological models (Kodama); the metric 
representation (Glikman-Kowalski); cylindrical waves quantization 
(Korotkin); bianchi type VII models (Manojlovic); spinors' 
evolution (Massarotti); and quantum cosmology (Barvinski).

The workshop was elegantly concluded by an inspiring talk by Jim 
Hartle, tiled ``Problems for the 21st century'': so that everybody, 
going home, could know what to do.  On another general relativity 
conference in Warsaw, half a century ago, Feynman wrote a famous 
comment, not too gentle towards the relativists.  Times are gone, and 
gravity is today a focal point of fundamental physics research.  Who 
knows, had he been there, maybe this time Feynman might have been a 
bit nicer \ldots
\vfill\pagebreak


\section*{\centerline {Alpbach summer school on fundamental 
physics in space}}
\addtocontents{toc}{\protect\medskip}
\addcontentsline{toc}{subsubsection}{\it 
Alpbach summer school on fundamental physics in space, by
Peter Bender}
\begin{center}
    Peter Bender, JILA, University of Colorado, Boulder\\
\htmladdnormallink{bender@jila.colorado.edu}{bender@jila.colorado.edu}
\end{center}
\parindent=0pt
\parskip=5pt

Each year a Summer School in an area of space science is held in the
picturesque mountain village of Alpbach, Austria.  Erwin Schroedinger
frequently spent time in the summer in Alpbach, and the main lecture
room in the Congress House there bears his name.  This year, the space
science subject chosen was Fundamental Physics in Space.  The school
was organized and supported by the Austrian Federal Ministry of
Science and Transport, the Austrian Space Agency, the European Space
Agency, the national space authorities of the ESA member states, and
the European Science Foundation.  The Chairman was Johannes Ortner
from the Austrian Space Agency.

About 50 graduate students from nine European countries took part in
the Summer School.  There were 25 lectures presented by scientists
interested in fundamental physics and in space.  The meeting started
with an opening talk by Roger Bonnet, the Science Programme Director
at ESA.  This was followed by a number of introductory talks covering
the early universe, gravitational physics, and the questions that can
be addressed by fundamental physics missions.  Talks on expected
improvements in accelerometers and clocks needed for gravitational
physics tests in space also were included.

Four main missions were discussed in the remaining lectures.  Two are
approved missions that are scheduled for flight.  One is the Alpha
Magnetic Spectrometer for the detection of antimatter in space and the
search for dark matter.  It will fly on the Shuttle and on the
International Space Station.  The other is the Gravity Probe B
mission, that will measure relativistic dragging of inertial frames
due to the Earth's rotation, and also geodetic precession caused by
the Earth's mass.

The other proposed missions have been the subject of international
studies, but are not yet approved.  One is the Mini-STEP mission,
where STEP stands for Satellite Test of the Equivalence Principle.
The differential accelerations of pairs of concentric masses will be
compared with great precision to determine if the ratio of
gravitational to inertial mass is the same for different elements.
The other mission is the Laser Interferometer Space Antenna (LISA) for
gravitational wave studies.  It will inventory thousands of galactic
binaries containing compact stars, and look for signals from sources
out to cosmological distances that contain massive black holes.

A unique feature of the School is that the students spent about half
their daytime hours in workshops, where they studied and prepared
proposals for possible future missions.  The students broke up into
two teams, "coordinated" by Robin Laurance from the European Space
Technology Research Center in The Netherlands and Nicholas Lockerbie
from Strathclyde University in Glasgow.  Each team worked on its
mission proposals during the workshops and often for many hours at
night, using ten PCs to search for information and carry out their
studies.  The proposals were presented on the last day of the School
at a session chaired by Hans Balsiger, the current ESA Science Program
Committee chair.

One team chose to concentrate their efforts on a mission to study
MACHOs - Massive Compact Halo Objects.  Intensive ground-based
measurement programs have detected nearly 100 temporary brightenings
of stars in the galactic bulge or the Large Magellanic Cloud due to
dark objects passing between them and us causing gravitational
lensing.  The objective of the space mission would be to detect small
displacements of star images as well as brightenings using a 1 m
diameter telescope and advanced microchannel plate detectors similar
to those being developed for particle physics experiments.  The
relative timing of pulses from dozens of stars would be determined as
a star field was swept across the roughly 10,000 parallel strip
channels of the detector by a rotating mirror.

The other team presented studies of five missions, of which three were
developed into specific proposals.  One proposal was for adding the
capability to the Alpha Magnetic Spectrometer to convert neutralinos
into detectable gamma-rays with energies of 30 to 300 Gev.  If this
can be done without losing sensitivity for antimatter detection, it
would permit searches for neutralinos from the galactic center, an
important dark matter candidate.

The other two proposals were for missions designed to considerably
exceed the sensitivity of the LISA gravitational wave mission at
frequencies lower than and higher than LISA is optimized for.  The
"extra low frequency" mission would improve the sensitivity for
sources involving supermassive black holes such as those found in
active galactic nuclei, and also would improve observations of
non-compact galactic binaries.  The "medium frequency" mission would
have its best sensitivity at frequencies between the optimum
frequencies for LISA and for ground-based detectors.  It would use
multiple bounces between mirrors in the different spacecraft.  The
main objectives would be to permit observations of neutron star binary
coalescence much earlier than possible on the ground, and to look for
possible coalescence of few hundred solar mass black holes in dense
galactic nuclei.

The Proceedings of the Alpbach Summer School, including descriptions
of the missions studied in the workshops, will be published by ESA.

\vfill\pagebreak


\section*{\centerline {
MG8, an experimentalists' summary}}
\addtocontents{toc}{\protect\medskip}
\addcontentsline{toc}{subsubsection}{\it 
MG8, an experimentalists' summary, by Riley Newman and Peter Saulson}
\begin{center}
    Riley Newman, UC Irvine and Peter Saulson, Syracuse University\\
\htmladdnormallink{rdnewman@uci.edu}{rdnewman@uci.edu} - 
\htmladdnormallink{saulson@suhep.phy.syr.edu}{saulson@suhep.phy.syr.edu}
\end{center}
\parindent=0pt
\parskip=5pt

   Here is an idiosyncratic selection of highlights of the 
experimental sessions of the 8th Marcel Grossmann meeting, held in 
Jerusalem 23-27 June 1997.

   Ken Nordtvedt, speaking as chair of a session on gravitational
experiments in space, gave a graceful review. He focused on the
aspect of experimental  gravity  that consists of the search for
(new) long range fields. This he placed in  the  context of the
general paradigm of physics that all interactions are carried by 
bosonic fields. From this point of view, the questions facing
experimental   gravity include the possible existence of non-linear
gravity, or of scalar, vector, or tensor fields in excess of those
included in general relativity.  Nordtvedt then reminded his listeners
of the tremendous success Lunar  Laser Ranging has had since the
Apollo 11 astronauts installed the retroreflector array on the
Moon. It is now fully competitive with  laboratory experiments as a
test of the Equivalence Principle, and is expected to keep pace with
improvements of lab experiments to reach  sensitivity to possible
variations in free-fall at the $1$ part in $10^{13}$ level. Excellent
prospects for future improvements in our empirical  knowledge could
come from two new classes of experiments: high-precision clocks
carried through the solar system (especially to the vicinity of the
Sun), and laser ranging (instead of radar ranging) to the
planets. Coupled with further studies of binary pulsars, Nordtvedt
predicted a long life ahead for this branch of experimental gravity.

   Experiments to observe the Lense-Thirring effect (dragging of
inertial frames by rotating masses) were discussed at several events
during the meeting. Ignazio Ciufolini described what could be done
with the series of LAGEOS satellites, dense spherical bodies studded
with hundreds of corner-cube reflectors that have been placed in high
Earth orbit. LAGEOS I and II are already in orbit, with a proposed
launch of LAGEOS III sometime in the near future. The L-T effect
should make the planes of the orbits of these satellites precess in a
characteristic way; Ciufolini has now claimed to have detected such
orbit precession at the $25\%$ level. Classical effects from the
non-sphericity of the Earth also cause precession, so the claim for
the detection of the relativistic effect rests on the assertion that
these less interesting effects can be accurately modeled. A lively
discussion among the attendees was dominated by a sense of optimism
that such modeling could be done well. A plenary talk by Francis
Everitt described progress on Gravity Probe-B, by all accounts to be
the definitive test of the Lense-Thirring effect. The experiment,
originally proposed by Leonard Schiff, involves a slightly different
version of frame dragging.  GP-B will carry four or five gyroscopes of
unprecedented symmetry, the precession of whose spin axes will reveal
the dragging of inertial frames.  The experimental plan includes a
rich mix of diagnostic tests that should give one confidence that
precession is due to relativity, and not to unmodelled classical
effects. The satellite is now making rapid progress toward completion,
with a launch expected some time in the interval Dec 1999 to Oct 2000.

   A number of interesting papers were presented at the session on
experimental tests of gravity, chaired by Cliff Will. Progress reports
on G measurements were presented by three groups.  The Wuppertal group
has increased its earlier estimates of uncertainty in field mass
positioning, and now reports $G =(6.6637 \pm 0.0004 \pm
0.0044)\times 10^{-11}$.  Their work continues, with a goal of a
$50-100$ ppm measurement.  The Z\"urich group, which will measure G by
measuring the weight changes of $1$ kg masses induced by steel tanks
containing $13.5$ tons of mercury, reported preliminary results using
water instead of mercury: $G=(6.674 \pm 0.001 \pm ?)\times 10^{-11}$,
with systematic error yet to be determined (currently estimated to be
$<600$ ppm).  The goal of the group is a $10$ ppm measurement.  The UC
Irvine group, which plans a G measurement with a cryogenic torsion
balance using a dynamic ("time of swing") method, reported
measurements of the properties of torsion fibers at low temperature
suggesting that anelastic fiber properties should not limit the
accuracy of such a G measurement at a $10$ ppm level or better. Also
presented were a $1/r^2$ test (A. Arnsek and A. Cadez) indicating that
the ratio of gravitational forces at distances of $30$ cm and $100$
cm agrees with Newton to about $1$ part/thousand, and a progress
report on the TIFR equivalence principle experiment, which anticipates
sensitivity to $\eta={\Delta a\over a}$ at a level $10^{-12}$ next year
and $10^{-13}$ in the future.  H.J. Paik described plans for a test
for ${\sigma}\cdot r$ dependent forces such as could be generated by an
axion, using a superconducting differential angular accelerometer with
target sensitivity more than five orders of magnitude greater than
current limits.  New space-based equivalence principle tests were
suggested by A. Nobili, who suggests that an $\eta$ sensitivity of
$10^{-17}$ may be achieved with a spring-tethered test mass system
rotating with its capsule at $5$ Hz, and by B. Lange, who proposes a
system of unconstrained concentric spherical shells in a drag-free
satellite.  Several talks suggested new types of EP tests in the realm
of atomic physics.  Ken Nordtvedt discussed GR tests that may be made
using clocks in solar orbit or a solar probe where redshift
measurements can be made in fields $U/c^2$ much larger than achieved
to date, with some scenarios suggesting sensitivity to $\gamma-1$
at a level as small as $10^{-6}$ or $10^{-7}$, to $\beta-1$ below
$10^{-6}$, as well as great sensitivity to the solar
J2 and possible EP violation in the form of different rates for clocks with
different dependencies on $\alpha$ .

   A special Memorial Symposium was held in honor of Robert H. Dicke, who
passed away in March of this year. Ken Nordtvedt spoke on Dicke's
thinking about Mach's Principle, particularly on whether general relativity
sufficiently embodies Mach's idea or instead if something like Dicke's
scalar-tensor theory is truer to Mach's vision. Symposium organizer Clifford
Will gave an overview of the key experiments carried out during Dicke's 
long career, including his many contributions to microwave physics and
astronomy, his improved Eotvos experiment, his early championing of
Lunar Laser Ranging, and his measurement of the solar oblateness. Brandon
Carter paid tribute to Dicke's proposal of the key idea that became known
as the Anthropic Principle. Francis Everitt spoke movingly of the 
inspiration he had drawn throughout his own career from the work of Dicke,
especially the new Eotvos experiment, as reported both in a preliminary
account in Scientific American and in the great 1962 treatise by Roll, 
Krotkov, and Dicke. He also reminded those in attendance of the influence
of Dicke's informal discussion group on gravitational physics at Princeton;
in 1957 one of its attendees was a Maryland physicist on sabbatical, Joseph
Weber. The session was rounded out by impromptu tributes from R. Cowsik,
H.J. Paik, and P. Saulson.

   A generous portion of time was allotted to work on the detection of 
gravitational waves, including sessions of contributed talks on resonant 
mass detectors, interferometers, and on calculations of waveforms from
astrophysics sources. There were also invited talks on various aspects of
the experiments given by Ken Strain (GEO and LISA), David Blair (UWA), and 
Piero Rapagnani (VIRGO).

   There were descriptions of several fascinating astrophysical phenomena
of obvious interest to relativists. Felix Mirabel gave a beautiful
review of the properties objects within our Galaxy that exhibit superluminal
motion (sometimes called "microquasars".) These objects appear to be
wonderful laboratories in which to test the Rees model of apparent
superluminal motion as an effect caused by light-travel-time effects when
emitting sources move at relativistic velocities in a direction not
parallel to the plane of the sky. Two review talks headlined a contributed
session on gamma ray bursts. David Band summarized the whole history and
phenomenology of the field since the first discovery of the mysterious events
in the 1970s. He was followed by E. Costa's outline of the new discoveries
made by the Italian satellite Beppo-SAX, whose multi-waveband 
instrumentation enabled observers to finally find optical, radio, and X-ray
counterparts to the enigmatic sources of the bursts. Now that the
cosmological distribution of these objects is apparently confirmed, 
modelers can focus their attention on the luminous end of the phase space
of models, most likely binary neutron stars that collide after spiralling
together due to gravitational radiation emission.
\vfill\pagebreak


\section*{\centerline {
Second Edoardo Amaldi Conference on Gravitational Waves}}
\addtocontents{toc}{\protect\medskip}
\addcontentsline{toc}{subsubsection}{\it 
Amaldi Conference on Gravitational Waves, by M. Alessandra Papa}
\begin{center}
    M. Alessandra Papa, University of Rome\\               
\htmladdnormallink{Maria.Alessandra.Papa@roma1.infn.it}
{Maria.Alessandra.Papa@roma1.infn.it}
\end{center}
\parindent=0pt
\parskip=5pt

The second edition of the Amaldi Conference on Gravitational Waves was
held in CERN this year from July 1 to July 4 and brought
together more than 150 scientists from 13 different countries.  Both
experimental and theoretical activity were well represented at the
conference and great emphasis was put on issues in data analysis.  The
programme included plenary talks, contributed communications and a
large poster session (27 posters), the latter two divided in five
topic workshops: sources, instrumentation, non-gaussian noise sources,
data analysis and future.

In the past decade a lot of effort has been spent by the resonant bar
detector groups to improve sensitivity and duty cycle of their
instruments. So, a great success, and in fact one of the highlights of
the conference, is the fact that presently there are {\it{five
gravitational wave detectors}} - all bar antennas - {\it{in continuous
operation}} in the world: NIOBE, in operation since 1993 at the
university of Perth, ALLEGRO, operating since 1991 (with a stop during
'95) at Louisiana State University, EXPLORER, taking data since 1990
(with a stop from '92 to '94, apart for a few months during '93), at
CERN, NAUTILUS in operation since 1996 at LNF (Frascati, Italy) and
the AURIGA detector, at LNL (Legnaro, Italy). The latter, as announced
during the conference, had started its first cryogenic run in february
'97 and has been in stable operation since then with a best
sensitivity around $8$ mK.
During the conference an agreement was signed among these groups to 
exchange data regularly on the basis of a common protocol.


The state of the art regarding the construction of the km-sized interferometric
antennas projects VIRGO and LIGO, and of the smaller scale
interferometers, TAMA 300 and GE0 600 was also reviewed. 
The schedules of all these projects foresee initial operation by the
year 2000. 



It clearly emerged that a great deal of effort is being made to
predict and model gw signals from astrophysical sources, especially 
black holes (W.H. Lee, R. Price, C.Palomba). 
B.S. Sathyaprakash showed that it is
possible to approximate wave forms of signals from inspiraling 
compact binaries so that they overlap with the
exact wave-form more than $95.6\%$ thus enhancing 
the detection probability to more than $90\%$. 
Issues regarding signals from 
{\it{binary systems}}, {\it{isolated stars}} and 
{\it{stochastic background radiation}} were addressed. 
G.Schaefer showed how to compute the secular changes of the orbital
parameters of a binary system up to order $1\over c^{10}$ thanks to ad
hoc balance equations between far zone fluxes and near zone losses, 
A.F.Zahkarov presented estimates of  $h\sim 8\times 10^{20}$ with
characteristic frequency at 1 kHz from $R\sim 50$ kpc  
for gw emission during non spherical
evolution of pre-SN in the framework of PN formalism. The 
results of E.Mueller from a comprehensive 
study of asymmetric core collapse
supernovae predict $h \le 4\times 10^{-23}$ 
for a source at $R\sim 10$ Mpc. 
M. Gasperini
presented predictions on a gravitational wave background
from the pre big-bang phase typical of string
cosmological model. At frequencies above 1 Hz, and up to 
about $10^{10}$ Hz, the expected
spectrum lies orders of magnitude (even 10) above that
predicted by standard inflation. Upper limits set by data from
detectors are still far from constraining the 
parameters of the model: the most recent data yield  
$\Omega_{gw}h^2_{100} \le 60$ and come from the cross correlation of 
EXPLORER and NAUTILUS data, whereas the upper border of the predicted 
value is at 
$\Omega_{gw}h^2_{100}\simeq 10^{-6}$. Nevertheless the future is
promising because already by cross correlating NAUTILUS, EXPLORER and 
AURIGA the upper limit could be lowered to $\Omega_{gw}h^2_{100}\simeq
10^{-3}-10^{-5}$.


In the data analysis session various topics were discussed. 
Hierarchical procedures to overcome the  
demanding requirements on computing resources needed to apply optimal 
matched filtering to the {\it  {search for unknown pulsar signals}}, 
have been presented by X.Grave and P. Astone. There have also been a
number of presentations (I.M.Pinto, A. Vecchio) 
on algorithms to estimate {\it {coalescing binaries}} 
parameters, both for space and ground based experiments.
A general point about what statistical approach, if bayesian or frequentist, 
should be used in gravitational wave data analysis was made by  
S. Finn in his talk. 

Future plans concern both the resonant mass and interferometric 
detection strategy. In the former category fall the projects for big
{\it {spherical detectors}}, of enhanced sensitivity and 
capable of estimating parameters of the
incoming radiation (E.Coccia, J.A. Lobo, S. Merkowitz ). For detection
at high frequencies ($\ge 2 kHz$) a {{\it local array}} of small multi mode
cylinders has been proposed. S. Frasca has presented data analysis
strategies for this instrument. In the latter category there is the 
space bound interferometer {\it{ LISA}} that could make observations
in the $ 10^{-4}-10^{-1}$ frequency range for signals from massive
black holes and galactic binary stars (J.Hough). 
\vfill\pagebreak


\section*{\centerline {
Santa Fe Workshop:} \\ \centerline{New Directions in Simplicial
Quantum Gravity}}
\addtocontents{toc}{\protect\medskip}
\addcontentsline{toc}{subsubsection}{\it 
Santa Fe workshop on simplicial quantum gravity, by Lee Smolin}
\vspace{-1mm}
\begin{center}
    Lee Smolin,	Penn State\\
\htmladdnormallink{smolin@phys.psu.edu}{smolin@phys.psu.edu}
\end{center}
\parindent=0pt
\parskip=4pt
\vspace{-2mm}
Much of the most interesting recent progress in quantum gravity concerns
approaches that are fundamentally discrete, in the sense that it is
assumed from the beginning that space or spacetime is built up out of
discrete structures.  This summer about 40 physicists working on a variety
of such approaches met for two weeks at Saint John's College in Santa Fe
to discuss recent progress in these areas.  Among the directions
that were represented were dynamical triangulations, random
surface theory, Regge calculus,
causal sets, decoherent histories, topological quantum field theory
and lattice and path integral approaches to non-perturbative quantum gravity.

The workshop was sponsored by Los Alamos National Laboratory and organized
by Emil Mottola.  The structure was informal and allowed much time for
discussions that probed the key issues in these areas.
Here is a summary of some of the highlights of the meeting.
(for more details as well as names and references I refer the interested
reader to the conference web site,
\htmladdnormallink{http://nqcd.lanl.gov/people/emil/sgrav.html}
{http://nqcd.lanl.gov/people/emil/sgrav.html}.

-Two dimensional random surface theory seems by now to be very well
understood.  The situation with four dimensional dynamical triangulations
is better, and the physics of the different phases is
better understood.  But the order of the phase transition is still
debated, although most participants seemed convinced by recent numerical
evidence favoring a weakly first order transition.  This led to lively
discussion as the standard scenario would imply that only theories with
a first order transition may have a continuum limit.  However, there
were proposals that theories with first order transitions may still
have critical behavior.  Another possibility discussed was that a
second order critical point might be found by varying a parameter associated
with the measure of the theory.

-There was lively discussion about the longstanding issue of the relationship
between Regge calculus and dynamical triangulations. Unfortunately, most of
the main proponents of the Regge calculus approach were absent, so a real
resolution was not possible.  However, it is clear there has been progress
on the issue of the measure of the path integral in Regge calculus.

-There are new and apparently very useful techniques for applying the
renormalization group to dynamical triangulations. 

-There has recently been a lot of progress in the causal set program.
One new idea is that directed percolation models may give examples
of causal sets which naturally have low spatial dimension.  These make
possible a new interchange with statistical physics in which 
methods from the study of directed percolation and cellular autonota
may be applied to elucidate non-perturbative behavior in quantum gravity.

-There are new connections between canonical quantum gravity, causal
sets, triangulations and topological quantum field theory.

-Analytical techniques may be applied to quantum gravity to uncover
the physics of the infrared behavior.  Under certain assumptions this
leads to surprising predictions about gravitation at cosmological
distance scales.  These and other analytical calculations may be compared
with the results of numerical simulations, leading to a very healthy
interaction of computational and analytical methods.
\vfill\pagebreak

\section*{\centerline {
VII Canadian Conference}\\\centerline{
on General Relativity and Relativistic Astrophysics}}
\addtocontents{toc}{\protect\medskip}
\addcontentsline{toc}{subsubsection}{\it 
VII Canadian Conference on General Relativity, by David Hobill}
\begin{center}
David Hobill, University of Calgary\\
\htmladdnormallink{hobill@crag.ucalgary.ca}{hobill@crag.ucalgary.ca}
\end{center}
\parindent=0pt
\parskip=4pt

This bi-annual conference took place at the University of Calgary on
June 5-7.  During the three day conference a total of nine invited
talks, thirty-five contributed talks and eight posters were
presented. In addition a special session was held in the memory of Ken
Dunn whose untimely death earlier in the year was met with great
sadness in the Canadian relativity community. Ken Dunn helped initiate
this series of conferences which began in 1985 at Dalhousie University
in Halifax. During the session held in his memory three talks were
presented by Jeff Williams, Tina Harriott, and Eric Woolgar (three
people closely with associated Ken).  All three talks were devoted to
recent results obtained from research on ``relativistic kinks''.

The invited talks covered a number of different topics. George
F. R. Ellis opened the conference with the first invited talk which
covered two different aspects of inhomogeneous cosmological
models. While issues regarding the Sachs-Wolfe effect and measurement
of the Cosmological Background Radiation were of great interest, a
lively discussion was generated by a new proposal for a definition of
gravitational entropy.

Other invited talks dealing with cosmological subjects were presented
by Bernard Carr who reviewed the status of various self-similar
solutions that might represent over- and under-dense compact regions
in the Universe and by John Wainwright whose talk was devoted to a
review of the evolution of the Bianchi Cosmological models and the
extent to which they undergo isotropization.

Black holes (and once again gravitational entropy and self-similar
solutions) were the topic of discussion by some of the other plenary
speakers. Richard Price discussed some recent results that have been
obtained using analytic approximation methods to compute the dynamics
of axi-symmetric black hole collisions.  Jack Gegenberg spoke on
gravitational solitons and how they may be used to represent black
hole spacetimes and Valeri Frolov presented a model demonstrating how
black hole entropy is generated in Sakharov's theory of induced
gravity. In addition a complete review of the status of research on
critical phenomena in gravitational collapse was presented by Matt
Choptuik.

On the observational/experimental side of general relativity Bruce
Allen provided a review of the latest results from, and progress being
made on the LIGO project, including an overview of the possible
sources.  In addition, Carol Christian of the Space Telescope Science
Institute presented a number of impressive images from the Hubble
Telescope and discussed how they have added to our understanding of
the universe around us.

Partial financial support for the conference was provided by the
Canadian Institute for Theoretical Astrophysics, the Fields Institute
for Research in Mathematical Science and the University of Calgary
whose generosity was much appreciated. Additional thanks go to Leroy
Little Bear (of the University of Lethbridge's Native American Studies
Department) who presented an interesting perspective on aboriginal
cosmological views during the conference banquet and to Big Rock
Brewery of Calgary who provided a special bottling of ``Black Hole
Ale'' which represented the first known industrial application of
black hole research.

\end{document}